\documentclass[12pt]{iopart}
\usepackage{iopams}
\usepackage {graphicx}
\usepackage {csquotes}
\usepackage{color}

\begin {document}

\title {In-flight measurements of energetic radiation from lightning and thunderclouds}

\author{Pavlo Kochkin$^1$, A P J van Deursen$^1$, Alte de Boer$^2$, Michiel Bardet$^2$, Jean-Fran\c{c}ois Boissin$^3$}

\address{$^1$ Department of Electrical Engineering, Eindhoven University of Technology, POBox.~513, NL-5600~MB Eindhoven, The Netherlands}
\address{$^2$ National Aerospace Laboratory NLR, Amsterdam, The Netherlands}
\address{$^3$ Airbus France, Toulouse, France}
\ead{p.kochkin@tue.nl, a.p.j.v.deursen@tue.nl}

\vspace{5cm}

{\color{red}Please cite the original paper on IOP Journal of Physics D: Applied Physics}

\maketitle

\begin{abstract}
In the certification procedure aircraft builders carry out so-called icing tests flights, where the zero degree Celsius altitude is deliberately sought and crossed in or under thunderstorms. Airbus also used these flights to test ILDAS, a system aimed to determine lightning severity and attachment points during flight from high speed data on the electric and magnetic field at the aircraft surface. We used this unique opportunity to enhance the ILDAS systems with two x-ray detectors coupled to high speed data recorders in an attempt to determine the x-rays produced by lightning in-situ, with synchronous determination of the lightning current distribution and electric field at the aircraft. Such data are of interest in a study of lightning physics. In addition, the data may provide clues to the x-ray dose for personnel and equipment during flights. The icing campaign ran in April 2014; in six flights we collected data of 61 lightning strikes on an Airbus test aircraft. In this communication we briefly describe ILDAS and present selected results on three strikes, two aircraft initiated and one intercepted. Most of the x-rays have been observed synchronous with initiating negative leader steps, and as bursts immediately preceding the current of the recoil process. Those processes include the return stroke. The bursts last one to four micro-second and attain x-ray energies up to 10 MeV. Intensity and spectral distribution of the x-rays and the association with the current distribution are discussed. ILDAS also continuously records x-rays at low resolution in time and amplitude.

\end{abstract}

\section{Introduction}
Thunderstorms and lightning are associated with high energy phenomena in several ways, as has been remarked in 1924 by Wilson \cite{Wilson1924}. Long gamma-ray glows from thunderclouds were first observed \cite{Parks1981,McCarthy1985} in a flight campaign carried out by a NASA F-106 jet. Later such glows were observed by balloons \cite{Eack1996,Eack2000} and from the ground \cite{Brunetti2000,Chubenko2000,Alexeenko2002,Torii2009,Tsuchiya2012,Chilingarian2010}. Their spectrum includes photons up to MeV energy range. The glows are a clear demonstration of Relativistic Runaway Electron Avalanches (RREAs) \cite{Babich1998} in nature.

On-ground observation of hard radiation from lightning began in 2001 \cite{Moore2001}. Further observations in natural and rocked-triggered lightning have been described in \cite{Dwyer2003,Dwyer2004,Dwyer2005b}; all known lightning leader types generated fast x-ray bursts with energies up to few hundred keV.

Terrestrial gamma rays flashes (TGFs) are intense bursts of extremely high-energy radiation which were first observed from space in 1994 \cite{Fishman1994} and which correlate in time with lightning discharges. The continuous energy spectrum of a TGF can reach 40~MeV according to \cite{Marisaldi2010} and perhaps even 100~MeV as remarked by \cite{Tavani2011}. It has been suggested that the downward intensity of a TGF poses a threat to aviation \cite{Tavani2013}. Because of safety requirements this needs to be clarified \cite{Trompier2014}.
Although the scales and energy ranges of these phenomena vary widely, their explanation starts from the same point - nonlinear behavior of electron friction force in ambient electric field (see for example \cite{Dwyer2004b}). The details of the mechanism are still much debated. Additional experimental data may help to unravel the problem. As an example, positron annihilation radiation has been detected by chance inside thunderstorms, as has recently been published \cite{Dwyer2015}. In general experimental data require flying or orbiting measurements setups; platforms should be exploited whenever available.

In this manuscript we describe an experiment to measure hard radiation from lightning and thunderclouds in situ. Aircraft manufacturer Airbus occasionally performs icing flights with test aircraft, also in the scope of certification of new aircraft. These prolonged flights at an altitude of zero degree Celsius occur in or under thunderstorms.
The Airbus 2014 icing campaign lasted 6 days. The aircraft has been struck by lightning 61 times, which is more than an average commercial airplane experiences during its entire lifetime.
The flights also give an opportunity to test and improve ILDAS, acronym for "In-flight Lightning Strike Damage Assessment System" \cite{ILDAS}. ILDAS was originally designed to determine the lightning current waveform and trajectory through an aircraft in-flight. We added the capability to measure the x-rays that are generated by lightning.
The x-ray data are registered with 10~ns time resolution during 1~second, well synchronized with the lightning strike data. At higher altitude the aircraft may trigger or intercept lightning. We present the first observations of x-rays inside the cloud, and demonstrate that x-rays occur during lightning leader initiation and at the attachment of a stroke to the aircraft.
The different phases of lightning attachment have been investigated extensively earlier \cite{Mazur1992a,Uman2003,Laroche2012}, and will only briefly discussed here. In view of our earlier laboratory studies on hard radiation generated in long sparks \cite{Kochkin2012,Kochkin2015}, we expected x-rays to occur at the stroke attachment to the aircraft.

First we discuss measurements at different altitudes with cosmic rays background only. The data are consistent with those measured in 1910's \cite{Kolhorster1913,Carlson2012}, and can even be inverted to provide the altitude. Then we select two lightning strikes initiated by the airplane and show them in highest time resolution available.
We also present one lightning strike intercepted by the aircraft.
We identify the current pattern through the airplane applying a simple model to the ILDAS magnetic field data and provide a plausible scenario based on the location of the cathode, the place where the electrons leave the aircraft. X-rays are detected during the initiation and propagation phases; all are closely associated in time with stepped leaders and recoil processes.

\section{The ILDAS system}
ILDAS includes a number of magnetic (H) field sensors with on-board data recording, and on-ground data analysis after the flight. The single electric (E) sensor mainly serves triggering purposes. Earlier lightning instrumented aircraft were equipped with a larger number of electric field sensors \cite{Laroche2012}. The final goal of the ILDAS project is a commercial product intended for regular airline operations as a real-time lightning damage assessment system. ILDAS is designed to determine intensity and attachment points of the lightning current. This information reduces maintenance time on the aircraft when hit by lightning. The ILDAS system has earlier been developed in an EU FP6 project with many partners [EU FP6-030806]; it has been described in more detail in a series of contributions to the ICOLSE conferences \cite{Zwemmer2009,Boer2011,Boer2013}.

\begin{figure*}[ht]
\centering
\includegraphics[width=0.7\linewidth]{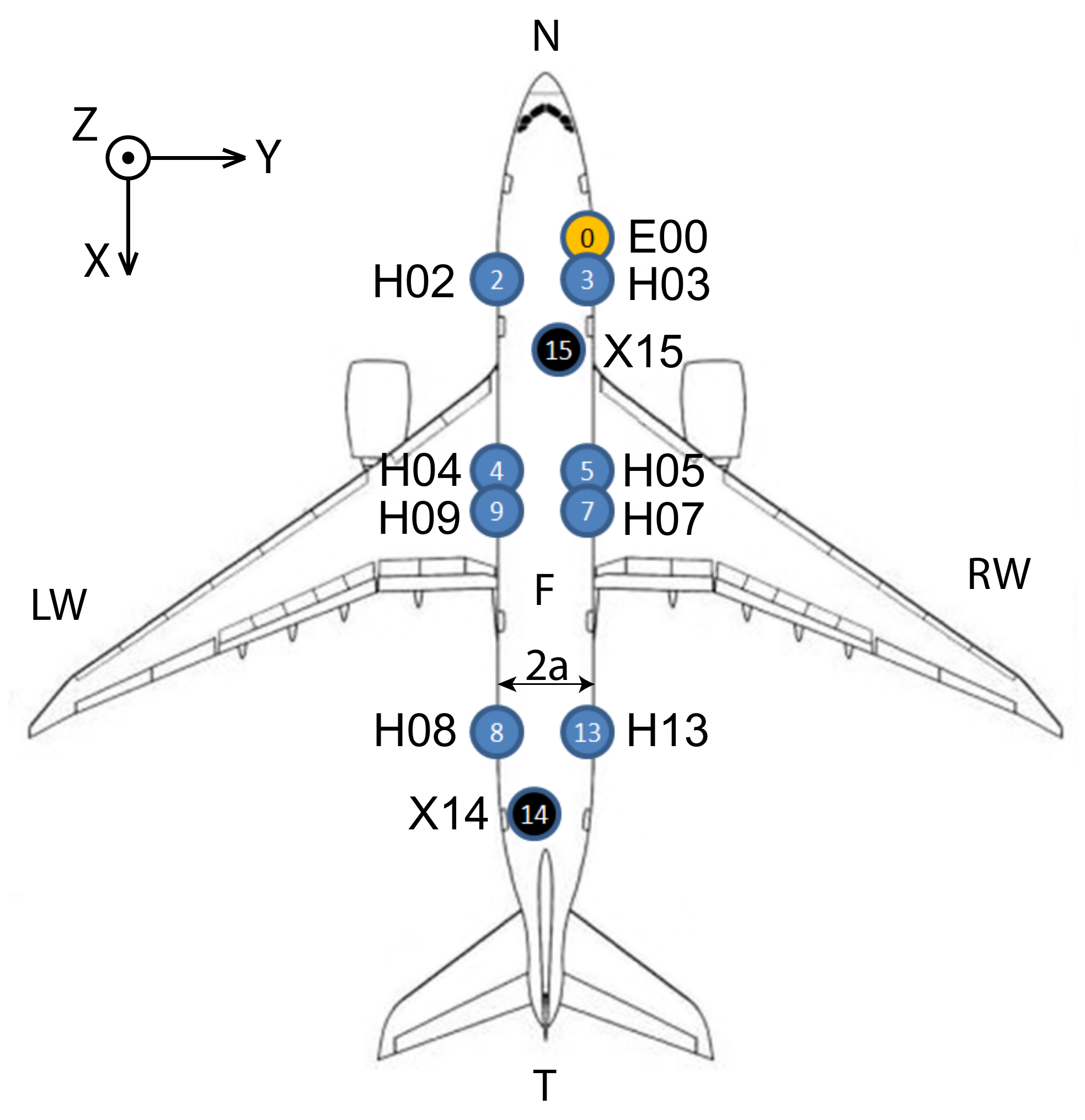}
\caption{ILDAS sensors location map for the A350 aircraft. The coordinate system referred to and the convention for left wing (LW), right wing (RW), nose (N), tail (T), fuselage (F) and diameter (2a) are indicated. Top-view of the aircraft is depicted.}
\label{fig:a350}
\end{figure*}

The lightning current through the aircraft induces a magnetic field of varying orientation and intensity over the surface. Inversely, when many simultaneous H-field measurements over the aircraft are available, one can determine the current pattern and attachment points. In a detailed electromagnetic model of the aircraft, the surface field has been calculated for a large number of pairs of current attachment points. The inversion follows from the best fit between calculated and measured H-field pattern. The method has been validated in an on-ground \cite{Zwemmer2009} measurement campaign on an A320 aircraft, employing a large set of sensors developed for ILDAS \cite{Stelmashuk2008,Zwemmer2009}. In the 2014 flights, Airbus employed only window mounted H-sensors which are described below. We do not use the detailed model, but instead derive sufficient information on the current pattern from the H-field polarities. Figure~\ref{fig:a350} shows the top view of the A350 aircraft used in our measurements. Its length is 66.6~m and the wingspan is 64.8~m. The position of the E and H-sensors and the x-ray detectors are indicated. All sensors and detectors are inside the temperature and pressure controlled cabin.

\subsection{ILDAS H and E sensors}
\begin{figure*}[ht]
\centering
\begin{minipage}{0.49\linewidth}
\includegraphics[width=\linewidth]{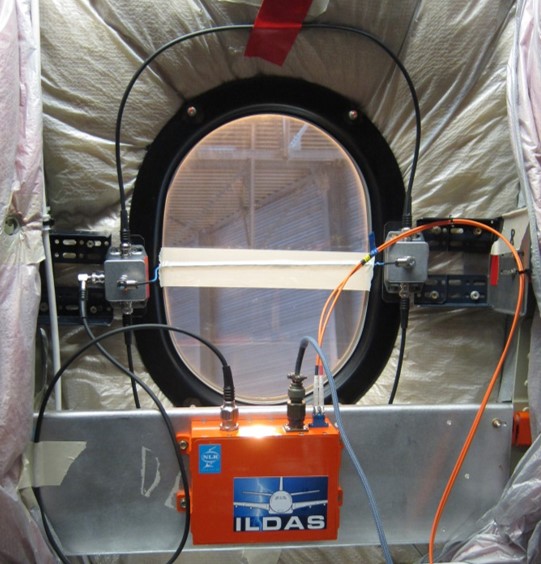}\end{minipage}
\begin{minipage}{0.49\linewidth}
\includegraphics[width=\linewidth]{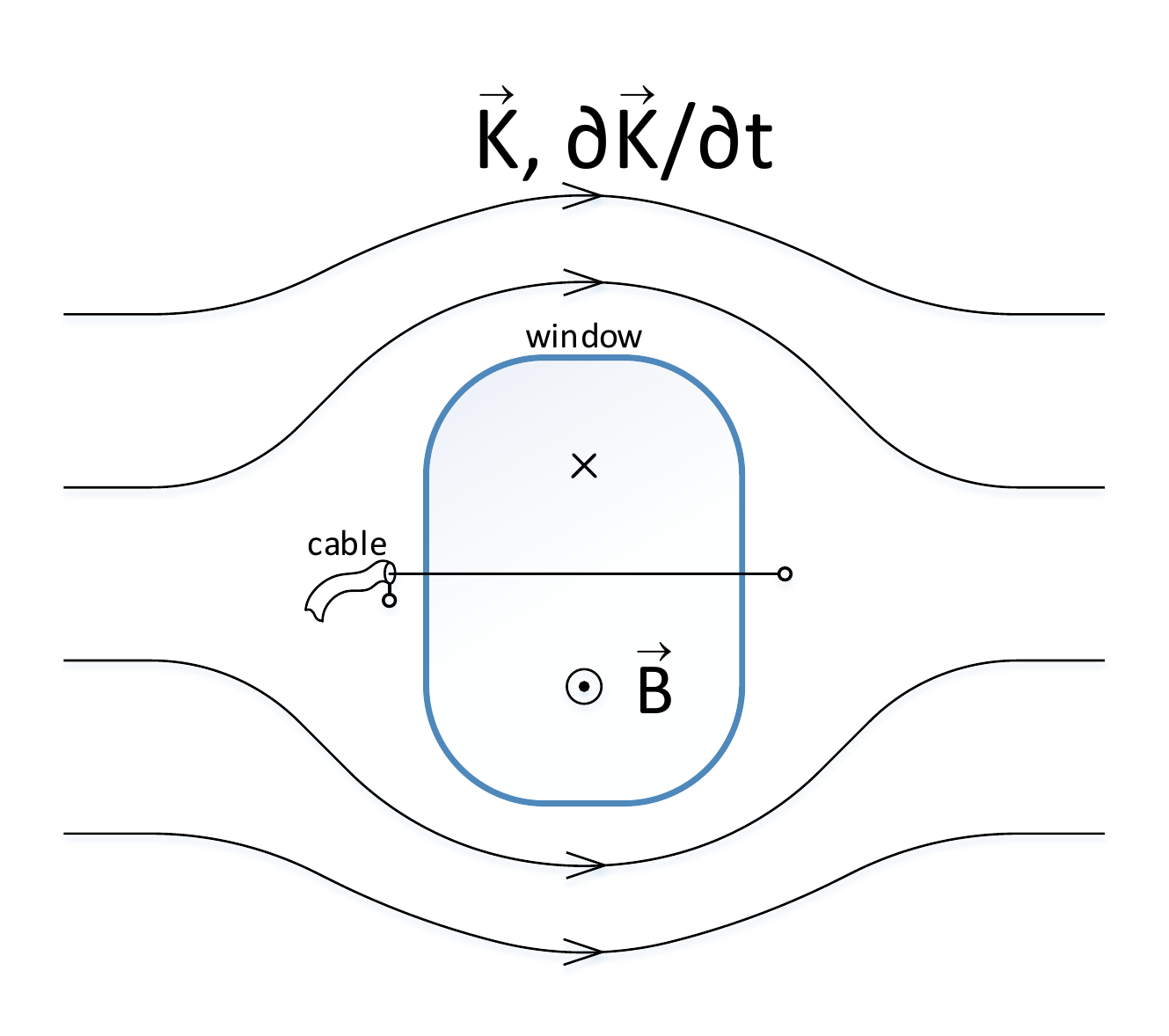}\end{minipage}
\caption{Photograph and drawing of the window sensor with sheet lightning current density $K$ through the aircraft fuselage. Magnetic field penetration is indicated by $\times$ and $\odot$. As shown, the sensor generates negative induced voltage for the indicated direction of the current density time derivative $\partial \vec{K}/\partial t $.}
\label{fig:window_sensor}
\end{figure*}

In a simple model without windows, the fuselage is a well conducting tube of radius~$a$. Assume a nose to tail lightning current $I$ homogeneously distributed over the circumference. Inside the fuselage the magnetic field is zero. On the outer fuselage surface one has a magnetic field $H_0 = I/2\pi a$, equal in magnitude to the sheet current density $K_0$; the vectors $\vec{H}_0$ and $\vec{K}_0$ are orthogonal. Windows modify the homogeneous current and magnetic field distribution as shown in figure~\ref{fig:window_sensor}, and can then act as H-sensor. We measure the voltage $V$ over the window at mid-height, for instance with a coaxial cable with the shield connected to the fuselage at one window side, and the inner conductor connected to the other side. The modified current density $\vec{K}$ causes some magnetic flux to enter and leave both window halves; see for instance \cite{Jackson1999} or \cite{Ollendorff1932, Kaden1959} for a circular window in a thin flat fuselage. The sensor sensitivity can be expressed in terms of an effective flux capturing area $A$ as $V = \mu_0 A \partial H_0/\partial t$, with $\mu_0$ the vacuum permeability. $A$ is of the same order as the window surface; the analytical solution for a circular window with radius $r_w$ is $A = r_w^2$. A numerical analysis of the window sensor takes the actual shape of the window and the mounting flanges into account \cite{vDeursen2011a,vDeursen2011b}. Measured and calculated values agreed better than 5\% for an aircraft with a thin metallic hull such as an Airbus~A320 \cite{vDeursen2011b}. The horizontal wire makes the sensor respond to the horizontal component of the current density $\vec{K}_0$; a vertical wire to the vertical component of $\vec{K}_0$, be it with different sensitivity because of the ellipsoidal window. With an aircraft radius $a$ of about 3~m, the conversion of H-field into net homogeneous fuselage current $I= 2 \pi a H_0$ becomes 2~kA per 100~A/m. Near the wings the conversion of magnetic field or current density into net current requires the detailed aircraft model. The photograph in figure~\ref{fig:window_sensor} shows a window sensor insulated from the fuselage; the loops close via two cables at some distance from the window perimeter rather than via the fuselage.

The sensitivity of the window sensor depends on the window material -- metal or composite.  For an A350 we assumed a metallic fuselage and window mounting pane \cite{vDeursen2013}. The calculation then slightly underestimates the actual effective area $A$ because the sensor misses the flux penetrating the composite parts.  A better but complicated model including the composite parts is in preparation. In addition, a part of the lightning current flows in the electrical structure network inside the aircraft with its proper magnetic field. If lightning current divides in an fuselage ($I_1$) and internal ($I_2$) part, each part contributes a magnetic field $H_{1,2}$ at the sensor. Then $K_0 =  H_1-H_2$ since $H_2$ appears at both sides of the fuselage; the sensors underestimate the total lightning current.

In the discussion that follows  a frame of reference with respect to the aircraft is useful. We follow the Airbus convention and assume the positive $x$-coordinate directed from nose to tail, the $y$-coordinate from left wing to right, and the $z$-coordinate vertically.
With the known orientation of the H-field sensors, we choose a positive plotted current signal corresponding to a $+x$ or $+y$ directed current.

\begin{figure}[ht]
\begin{minipage}{0.49\linewidth}
\includegraphics[width=\linewidth]{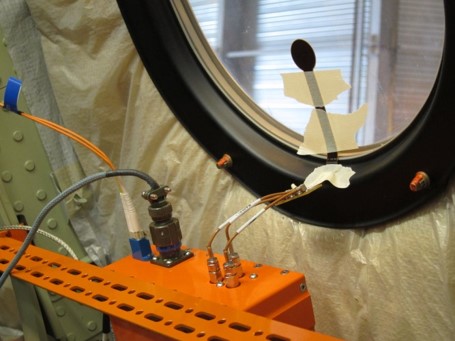}\end{minipage}
\begin{minipage}{0.49\linewidth}
\includegraphics[width=0.905\linewidth]{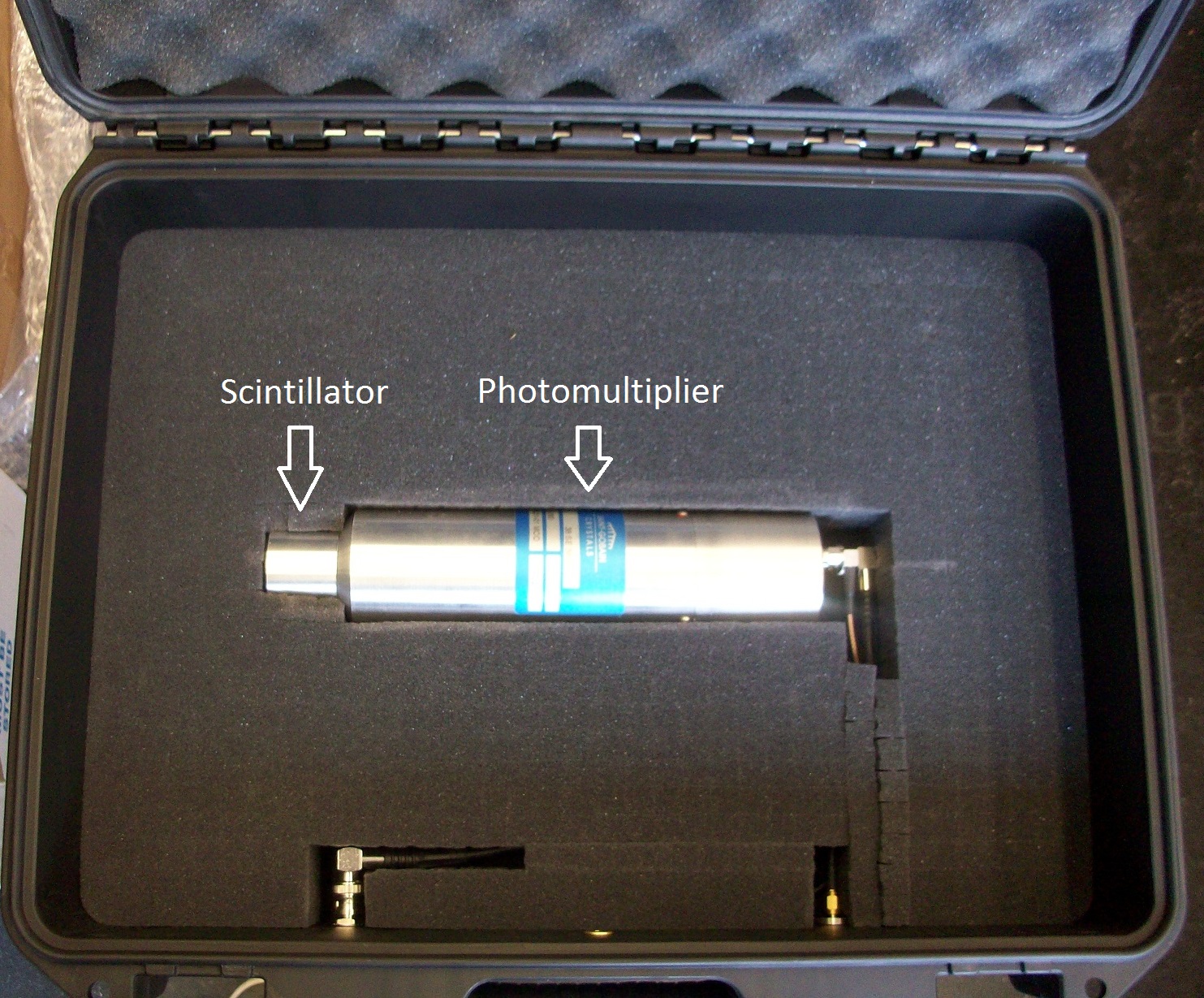}\end{minipage}

\caption{\textit{(Left)} The electric field sensor consists of a capacitive probe attached to a window and connected to its Sensor Assembly Electronics box (SAE). \textit{(Right)} One of the X-ray LaBr$_3$ scintillation detectors.}
\label{fig:EX-sensors}
\end{figure}

The electric field sensor consists of a capacitive probe attached to a window (see figure~\ref{fig:EX-sensors}). The E-sensor is commonly mounted near the window edge, which enhances the local electric field over to the near-to-homogenous field on the fuselage at some distance from the window. This effect has been included in the conversion \cite{Airbus2012}.
In the original ILDAS concept the E-field sensor only served as trigger source because the primary goal was the current distribution obtained from the magnetic field data. The chosen polarity of the E-field sensor is identical to \cite{Mazur1992a}: a positive signal corresponds to electric field lines pointing towards the aircraft, or towards a negative charge density on the fuselage near the sensor. The data presented below also include an additional correction factor of 2 due to the particular mounting of the sensor during these flights.

Both E- and H-sensors are differentiating sensors. In order to redress the frequency response, the first stage of the signal acquisition chain is  a combined passive/active analog integrator. ILDAS continuously monitors each signal in an separate Sensor Assembly Electronics (SAE), the orange box in figures~\ref{fig:window_sensor} and in \ref{fig:EX-sensors}). The sampling rate of the analog to digital converters (ADC) for the H and E sensors is 83.3~MS/s (12~ns sampling time). The dynamic range is 96~dB over the frequency band from 100~Hz up to 10~MHz for H-sensor and 10~Hz up to 500~kHz for E-sensor. The characteristic variation of the electric and magnetic field upon lightning attachment to the aircraft surface provides a trigger for ILDAS. After the trigger the systems stores pre- and post-trigger data over a time span of 1.2~seconds for the H and E sensors. In the process all SAEs are synchronized to within 1~sample at 5~ms intervals.

\subsection{X-ray detectors}\label{x-ray_detectors}
For our experiments the ILDAS system has been enhanced with two x-ray detectors, X14 and X15 in figure~\ref{fig:a350}. A photograph of a detector inside its shock protection box is shown in figure~\ref{fig:EX-sensors}. The detectors consist of LaBr$_3$(Ce) scintillators; we choose this material because of its quick response with 11/16 ns rise/fall time and good photon yield \cite{Dorenbos2004}. The crystals have a diameter of 38~mm and a length of 38~mm. The scintillation light is amplified by Hamamatsu 10-stage photomultiplier with special dynode voltage dividers to enhance the maximum acceptable pulse frequency \cite{SaintGobain}. The detectors, the calibration and the procedure for signal reconstruction have been discussed in \cite{NguyenCV2012} and \cite{Kochkin2015}.

\begin{figure*}[ht]
\centering
\includegraphics[width=0.7\linewidth]{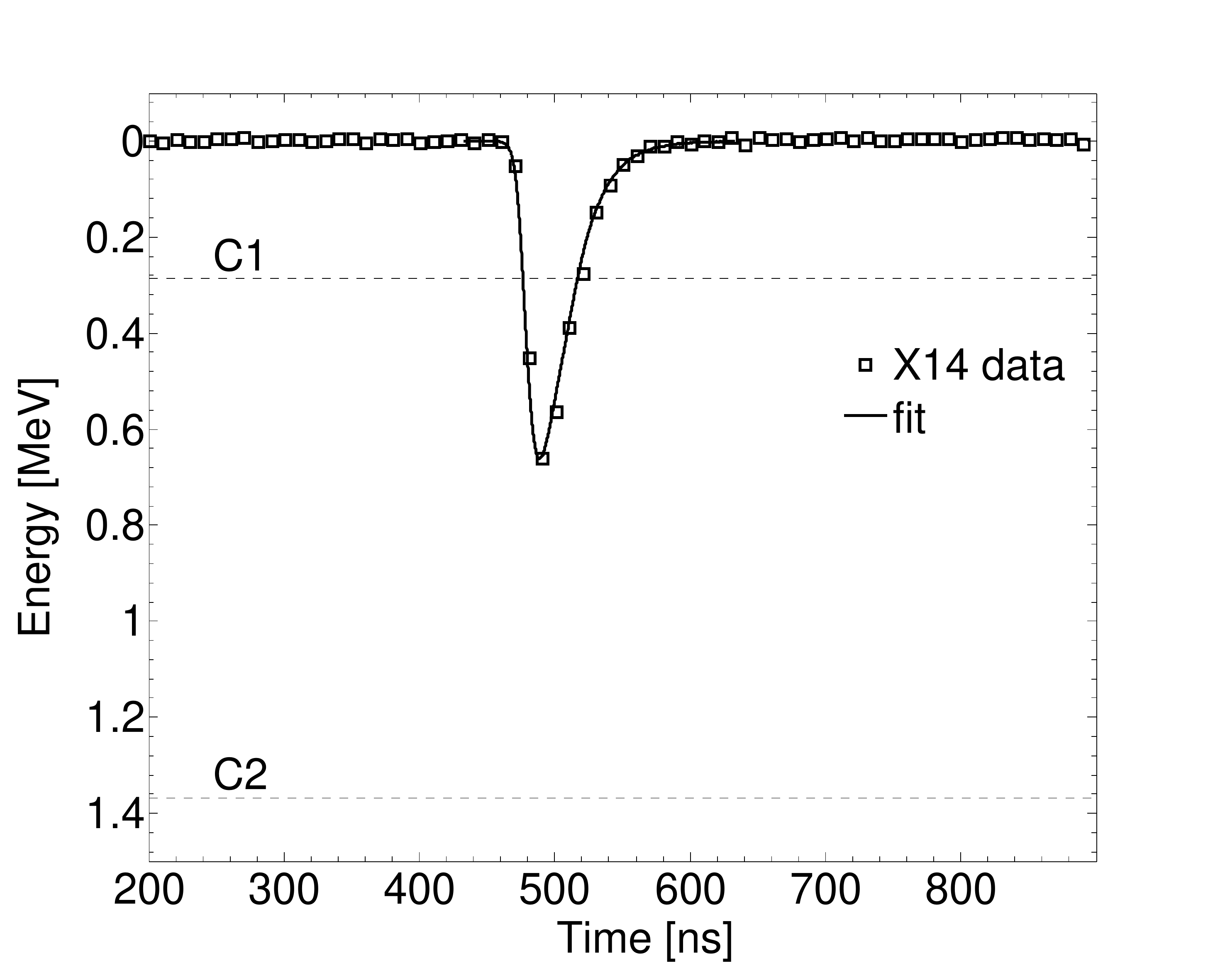}
\caption{A single 662~keV photon detection by X14 from $^{137}$Cs calibration source (squares). The solid black line is a single photon response function obtained by averaging many 662~keV photon signals. The horizontal dashed lines indicate C1 and C2 levels for continuous data file (CDF) record. For this single photon signal, the C1 counter increases by 4 (the number of squares below the level), while C2 remains constant.}
\label{fig:137Cs}
\end{figure*}

The SAEs for the X-ray sensors continuously record the analog photomultiplier output at 100~MS/s sampling frequency and store the data during 1~second upon triggering, again with some pre- and post-trigger time. The 100~Ms/s suffices to distinguish individual photomultiplier pulses and to discriminate real signals from interference.
As an example, in figure~\ref{fig:137Cs} the squares show the SAE data of a single photon detection by X14 from $^{137}$Cs calibration source. The solid black line represents the response when averaged over many single photon signals.

The aircraft fuselage, the protective box and the scintillator housing absorb x-rays with an energy of 30~keV and lower. As a result we only see the harder part of the x-ray spectrum. Cosmic rays cause a never-interrupted background of high energy x- and $\gamma$-rays and charged particles; therefore the x-ray detectors cannot assist in the triggering of ILDAS. This background also varies strongly with altitude.
In view of the long duration measurements obtained in earlier aircraft \cite{McCarthy1985,Parks1981} and balloon \cite{Eack1996,Eack2000} flights, it is desirable to also obtain information on x-rays activity outside the 1~second high-speed coverage. We implemented three special software counters in the SAE. The counters continuously determine the number of 10~ns sample periods where the x-ray signal exceeds three preset energy levels over consecutive periods of 15~ms. During flight the data are stored in continuous data files, CDF for short. The CDFs later provide information about the background and to a limited extent about its energy spectrum. The three preset levels for X14 are C1 = 0.29, C2 = 1.37 and C3 = 13.50~MeV; for X15 are C1 = 0.64, C2 = 1.96 and C3 = 15.30~MeV.
The dashed lines in  figure~\ref{fig:137Cs} indicate the C1 and C2 levels. This example pulse adds four to the C1 counter, and zero to C2 and C3.

\begin{figure*}[ht]
\centering
\includegraphics[width=\linewidth]{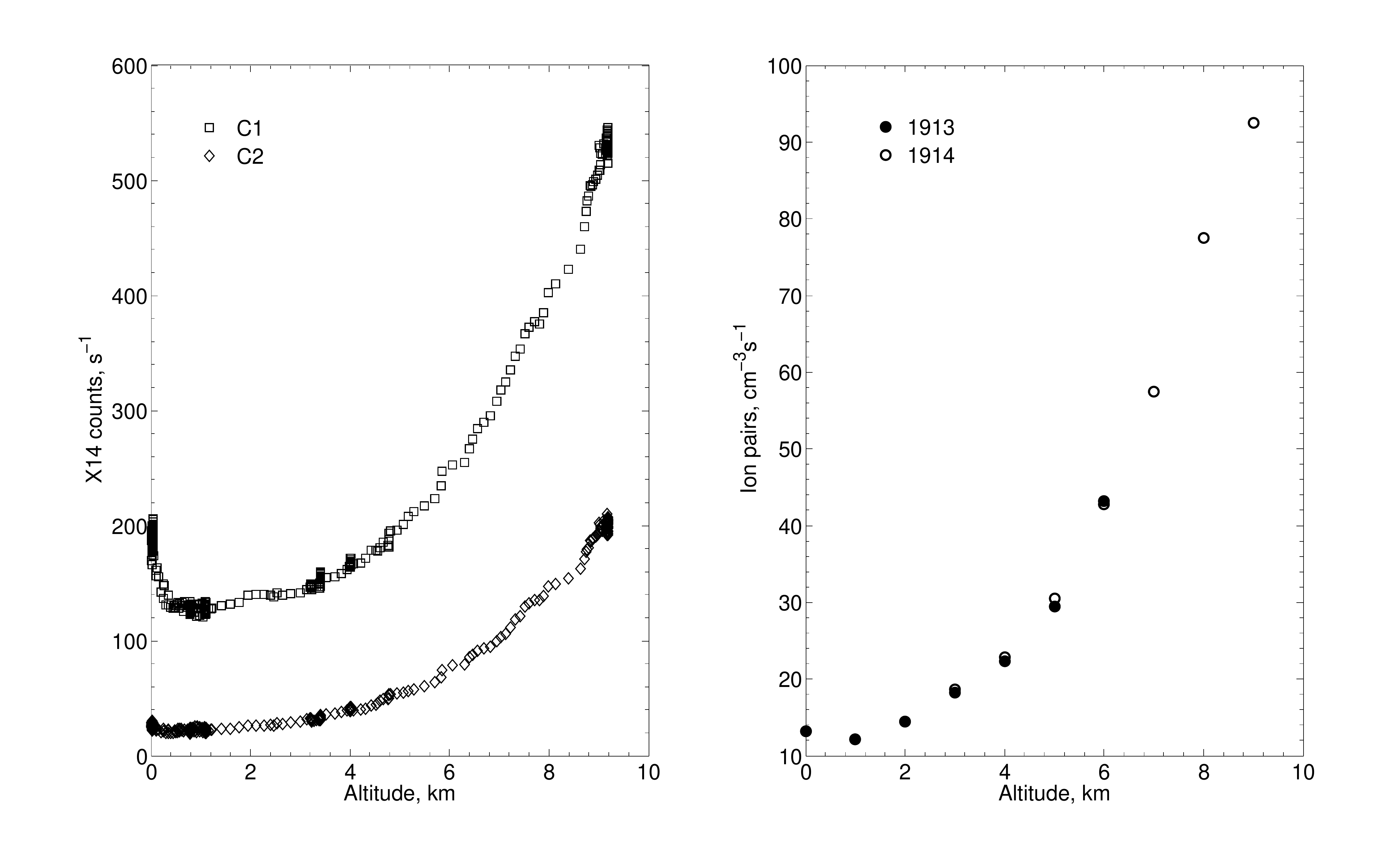}
\caption{\textit{(Left)} The number of 10~ns sample periods that the x-ray detector output exceeded the levels of 0.29~MeV (C1) and 1.37~MeV (C2). The data expressed in counts per second are the running average over 75~s. The flight occurred in December 2013. \textit{(Right)} The rate of atmospheric ionization as a function of altitude, as measured by Werner Kolh\"{o}ster in 1913-14 \cite{Kolhorster1913}.}
\label{fig:kolhorster}
\end{figure*}

The first test flight occurred in December~2013, when the ILDAS system with embedded x-rays detectors was flown in an A350 over a range of 300~km between Bergerac and Foix in Southern France. The flight lasted a little over 2~hours and was at 9~km altitude for 20~minutes. No thunderstorms were crossed, which allowed us to test the x-rays detectors on the cosmic ray background. Figure~\ref{fig:kolhorster} on the left shows the plot of the CDF data for detector X14, where we plotted the number of 10~ns sample periods that the detector output exceeded the levels of 0.29~MeV (C1) and 1.37~MeV (C2). The data are re-scaled as number of sample counts per second. In order to reduce noise we plotted the running average over 75~s. The horizontal axis is the aircraft altitude obtained from the aircraft log data. Since the X15 results are similar we did not included these in the plot. In the right part of the figure shows the original 1913-14 cosmic ray data measured by Kolh\"{o}rster \cite{Kolhorster1913,Carlson2012}. At ground level there is a strong contribution by radioisotopes emanating from the soil. At higher altitudes cosmic rays dominate. Both figures show a minimum around 1~km altitude. The similarity between the data is striking. The major difference occurs for the C1 data that either show an low energy excess at lower attitude, or a lack at higher altitudes. But one should also consider that we compare unequal quantities: counts of ADC sampling periods (left) versus number of ion pairs created (right).

\section{Continuous data file}
\begin{figure*}[ht]
\centering
\includegraphics[width=\linewidth]{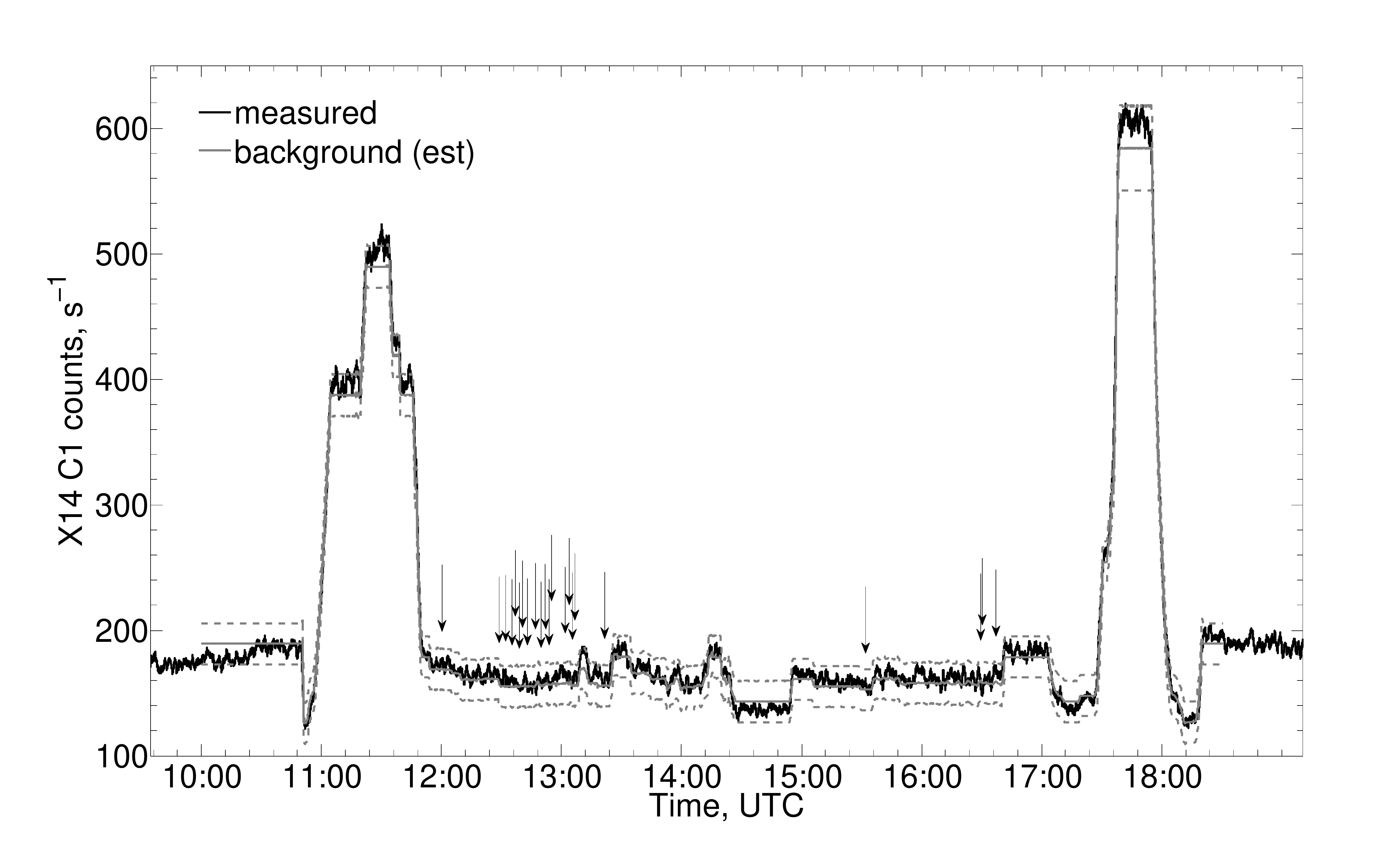}
\caption{Measured and estimated with 95\% confidence boundaries radiation inside the aircraft based on the continuous data file (CDF) content, recorded during the flight on April 30, 2014. The 22 downward arrows indicated lightning strikes. The background radiation inside the thundercloud (from 12:00 till 17:00) never exceeded the expected level.}
\label{fig:bgnd}
\end{figure*}

Figure~\ref{fig:bgnd} shows the continuous data file (CDF) data recorded on April~30, 2014 along with the estimated background level calculated from the flight altitude profile and the December 2013 data shown in figure~\ref{fig:kolhorster}(a). The airplane takes off at about 10:50, gains altitude and reaches a thundercloud at about 11:50. Then it descends to zero-degree level at about 4~km height and maintains that altitude with small variations. After 17:10 the aircraft gains altitude to fly back to the base and finally lands at about 18:20. The CDF data have been averaged over 75~sec as discussed above in section~\ref{x-ray_detectors}. Twenty-two lightning strikes were detected and recorded during this flight.
The strike times are shown by downward arrows. In the averaged data the lightning strikes are not clearly visible.
In view of the orders of magnitude intensity increase reported in \cite{Eack1996} we would expect a $\gamma$-ray glow to appear as an excess over the altitude derived background level outside the 95\% confidence band. No such excess has been found.

\section{Three selected lightning strikes}
The icing campaign of April 2014 consisted of six flights of an Airbus~A350. The aircraft takes off in Toulouse and heads to thunderstorm centers over southern Europe within an hour flight at cruising altitude. There the aircraft descends to the zero degree Celsius level which is at about 4~km during the season; the aircraft speed is then approximately 150~m/s. We selected two strikes classified as an aircraft-initiated lightning, and one aircraft-intercepted out of 61 recorded that contain most of the observed features \cite{Mazur1992a,Uman2003,Mazur1989}.

\subsection{Aircraft initiated strike 5049BB21}\label{5049BB21}
\begin{figure}[p]
\begin{minipage}{\linewidth}
\includegraphics[width=1\linewidth]{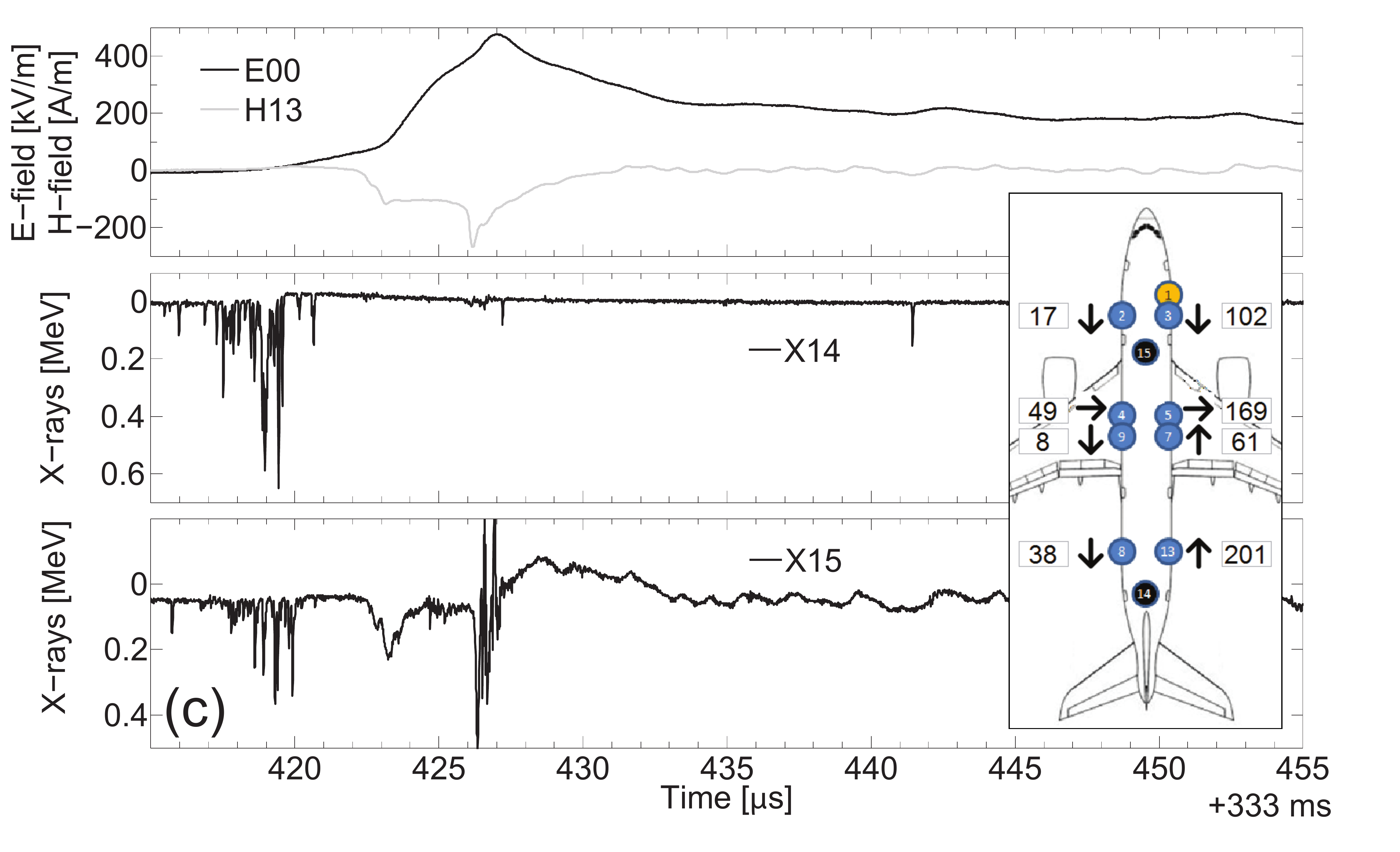}
\end{minipage}
\begin{minipage}{\linewidth}
\caption{(\textit{a}) Shows the data of the strike on the A350 recorded at 12:42:05~UTC on April~30, 2014, our event code 5049BB21. (\textit{b}) zooms in on the strike initiation, with two x-ray pulses, one clearly associated with a step in the E-field. The inset shows the current density distribution at the largest peak at $t=-4.5$~ms. In the inset the numbers give the sheet current density $K_0$ at the sensor position in A/m, and the arrows indicate the direction. (\textit{c}) zooms in on the single return stroke, and shows an simultaneous x-ray burst on both detectors; the bursts last about 4~$\mu$s. The inset shows the current density distribution at the peak of the return stroke at $t=426~\mu$s (+ 333~ms).}
\label{fig:5049BB21}
\end{minipage}
\end{figure}

Figure~\ref{fig:5049BB21}(a) shows the strike on the A350 recorded at 12:42:05~UTC on April~30, 2014, our event code 5049BB21. The E and H data come from sensor E00 and H13 respectively, the x-ray data from X14 and X15.
The high speed data record starts with the characteristic E-field variation for aircraft triggered lightning \cite{Laroche2012}, immediately followed by a burst of intense H-field spikes. Figure~\ref{fig:5049BB21}(b) zooms in on the initiation phase on a millisecond scale. The small letters between brackets in the lower part of Figure~\ref{fig:5049BB21}(a) show the times of the zoomed parts. For the E-field variations we follow the commonly accepted explanation described in \cite{Laroche2012,Mazur1989,Mazur1992a} and summarized in \cite{Uman2003}.

The aircraft-initiated discharge starts at $t = -9$~ms with a positive leader on an aircraft extremity, which usually happens when the ambient electric field is approximately 50~kV/m \cite{Uman2003}.
The E-field rises between markers A and B up to about 150~kV/m. The corresponding negative charge on the aircraft is of the order $-3$~mC. The current associated with this leader is outside the band and sensitivity limit of the window sensors. Between B and C, a negative stepped counter leader forms at the right wing. Between C and D the aircraft is a part of the current path and the E-field is not related to the total aircraft charge any more. The inset of figure~\ref{fig:5049BB21}(b) shows the current density pattern for the largest current step.
The current density (in A/m) near the fore sensor H02 is opposite to H03. This indicates a partial loop current over the fore part of the aircraft, induced by the wing-to-wing current. A similar opposition and loop is present at the aft sensors H08 and H13.

Left and right sides do not add to zero, due to a small net current on the aircraft near nose or tail, or to an off-vertical magnetic field component.
X14 (aft in the aircraft) detected two x-ray signals; both appear as single photon pulses. The larger one with 0.22~MeV near $t = -8$~ms is not related to any clear change in E or H on the aircraft. It is likely due to cosmic ray background. The smaller one with 0.09~MeV comes $0.3~\mu$s after the E-field step near B belonging to the negative leader.
In view of the smaller energy, it should originate within several tens of meters near the aircraft since otherwise it would have been absorbed by the air. No simultaneous signal has been observed on X15. During this strike we observed  56~x-ray photons in a second that could not be associated to a particular lightning feature in the E or H-data. The chance to detect such a background photon in 10~$\mu$s window around the E-field pulse is equal to 5.6$\cdot$10$^{-4}$.

A remarkable feature is the 0.33~s delay between the initiation and the first return stroke. Over this 0.33~s period there is only limited sensor signal both for E and H.
Over the 0.33~s time span the aircraft moved about 50~m, which led to the walk, sweeping or sudden jump of the attachment point from left wing to the tail. Figure~\ref{fig:5049BB21}(c) presents the signals of E00 and H13 zoomed-in on the return stroke on a microsecond scale; the time axis is shifted over 333~ms.
Between $418 < t < 422~\mu$s (+ 333~ms) there is some small initiation currents with most directions identical to the first initiation.
The return stroke peaks at $t=426~\mu$s (+ 333~ms) and has opposite current direction; compare the insets of figure~\ref{fig:5049BB21}(b) and (c). The signals of all H-field sensors combined indicate a dominant current path from tail to right wing.
Both detectors X14 and X15 recorded a large number of intense x-ray pulses during the $4~\mu$s long initiation of the return stroke. The pulses in the burst can be described by single photon responses of the detector.
The energy per pulse and the number of pulses increase with time for both detectors. We favor a single X-ray source approaching the aircraft.
In a tentative interpretation for the x-ray intensity, we note that the first initiation leads to a wing-to-wing current, which changes 0.33~s later and 50~m further into predominant tail-to-right-wing current with the tail acting as cathode. The main lightning channel remains stationary, while the aircraft moves. The attachment has to build up again in the jump from wing to tail. Large electric fields occur because of the leader charge, and electrons are sufficiently accelerated to cause x-rays by bremsstrahlung. During the $4~\mu$s initiation of this stroke, the tail acts as cathode.
In the laboratory \cite{Kochkin2012,Kochkin2015} we showed that the most probable x-ray source is located near the cathode.
This is consistent with the larger x-ray flux of X14 compared to X15 and the larger current density at H13 near the tail compared to the nose at H03.

The X15 detector picks up some interference caused by the magnetic field variation, as is apparent between 422 and 424~$\mu$s in Figure~\ref{fig:5049BB21}(c) and between 426 and 428~$\mu$s. The X15 detector is placed between the nose and the wing box where one may expect a concentration of the lightning current and enhanced magnetic fields.
Definitely, the interference on the X14 detector is less. Some modification of the x-ray detector - SAE combination is desirable, either in electromagnetic shielding or in position. All other X-ray pulses in figure~\ref{fig:5049BB21}(a) are single photons and can be attributed to background.

\subsection{Aircraft initiated strike 203C2BF0}\label{203C2BF0}
\begin{figure}[p]
\begin{minipage}{\linewidth}
\includegraphics[width=\linewidth]{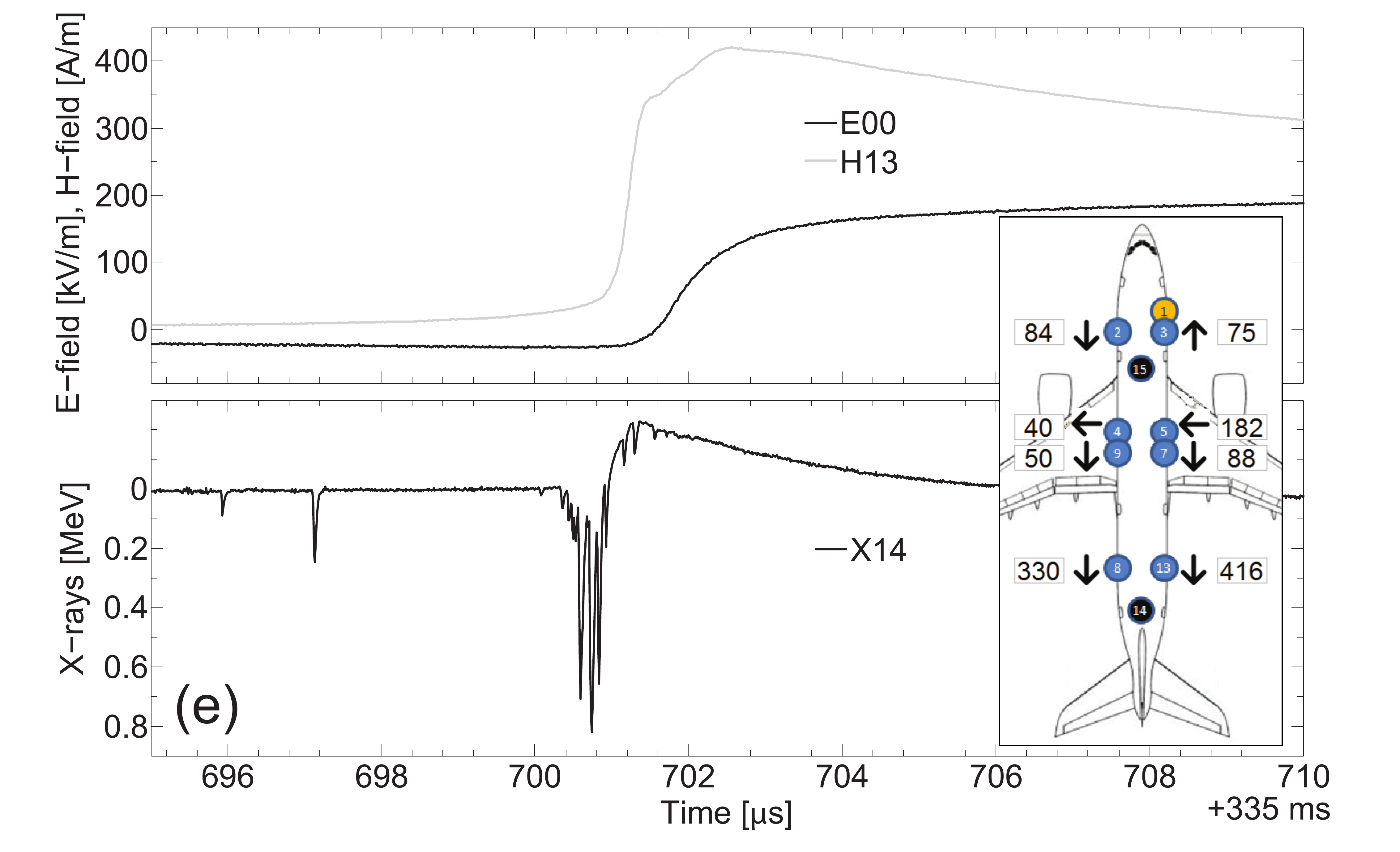}\end{minipage}
\begin{minipage}{\linewidth}
\caption{(\textit{a}) Shows 0.8~s of the measurements during the A350 flight on April~24, 2014. A lightning strike occurred at 16:22:20~UTC, our event code 203C2BF0. (\textit{b}) Zooms in on the attachment phase at $t=-80$~ms. The inset shows the sheet current density distribution at the largest peak. Each arrow in the inset shows the sheet current density direction, while the numbers indicate absolute current density in A/m. (\textit{c}) Zooms in on the 10~MeV x-ray pulse at $t=302$~ms. A more variable x-ray burst of the event of about 1~$\mu$s duration and lesser energy occur at $t=319$~ms (\textit{d}) and at $t = 335$~ms (\textit{e}).}
\label{fig:203C2BF0}
\end{minipage}
\end{figure}

Figure~\ref{fig:203C2BF0}(a) shows 0.8~s of the measurements during the A350 flight on April 24, 2014. A lightning strike occurred at 16:22:20 UTC, our event code 203C2BF0. The E and H data come from sensor E00 and H13 respectively, the x-ray data from X14. The X15 detector was not operational this flight.
Many current pulses are accompanied by x-rays. But not all x-ray signals match with a current pulse and may be due to background.

Figure~\ref{fig:203C2BF0}(b) zooms in on the discharge initiation at $t = -82$~ms. The E-field variations are similar to figure~\ref{fig:5049BB21}(b) and the letters A---D designate the same phases as before. Two single x-ray pulses of about 2~MeV appear during E-field steps, although the association is weak. The first x-ray photon occurs $13~\mu$s after the step in E, while the second occurs $2.3~\mu$s before the step in E. Between markers B and C one or more negative stepped leaders grow from the aircraft. The steps go with a current pulse of the order of 500~A.
All current pulses after 0.1~s are \emph{recoil processes} which is a collective name for the first or subsequent return strokes, dart leaders, recoil streamers or M-components \cite{Laroche2012}.

Figure~\ref{fig:203C2BF0}(c) zooms in on the stroke at $t=0.302~s$. The pattern shown in the inset suggests a 7~kA current from tail (cathode) to right wing. The strong current densities near the nose appear to be induced. On the rising edge of the current oscillations can be noticed at a frequency of approximately 3~MHz, which can be attributed to an electromagnetic mode of the aircraft. No such signature is seen in the E-field data because of the 500~kHz bandwidth limitation. An intense x-ray pulse occurs at the start of the current, simultaneous with the steepest current density rise and the largest negative E-field variation. The total x-ray burst lasts about $1~\mu$s. The signature of small individual x-ray pulses or photons can be discerned at the begin near 220~$\mu$s. The intense 10~MeV peak appears as a pile up of at least two comparable single photon pulses within 100~ns. However, such large x-ray energies lead to some saturation of the detector photomultiplier \cite{NguyenCV2012} and pulse widening.

Figure~\ref{fig:203C2BF0}(d) shows the more structured x-ray burst of about 1~$\mu$s duration at $t=0.319~s$. The double peak appearance suggests two overlapping components or events. The x-rays again occur just before the current sets in and near the largest E-field variation. The current attains a maximum of 5~kA, again flowing from tail to right wing.

Figure~\ref{fig:203C2BF0}(e) shows a similar burst of lesser x-ray energy, again occurring at the begin of the current pulse of 8~kA peak. Compared to the stroke shown in figure~\ref{fig:203C2BF0}(d) the current pattern reversed, while the electric field variation near the nose is positive in both cases, indicating a negative charge on the aircraft. In general, when we look at the full time span between $t = 0.22$ and 0.36~s, or about 20~m displacement of the aircraft, many strokes can be distinguished. The relative current pattern does not vary much. However, the current direction reverses at least four times. The x-ray activity appears less affected by the current reversals.

\subsection{Aircraft intercepted strike 504EC33C}\label{504EC33C}
\begin{figure}[ht]
\begin{minipage}{\linewidth}
\includegraphics[width=\linewidth]{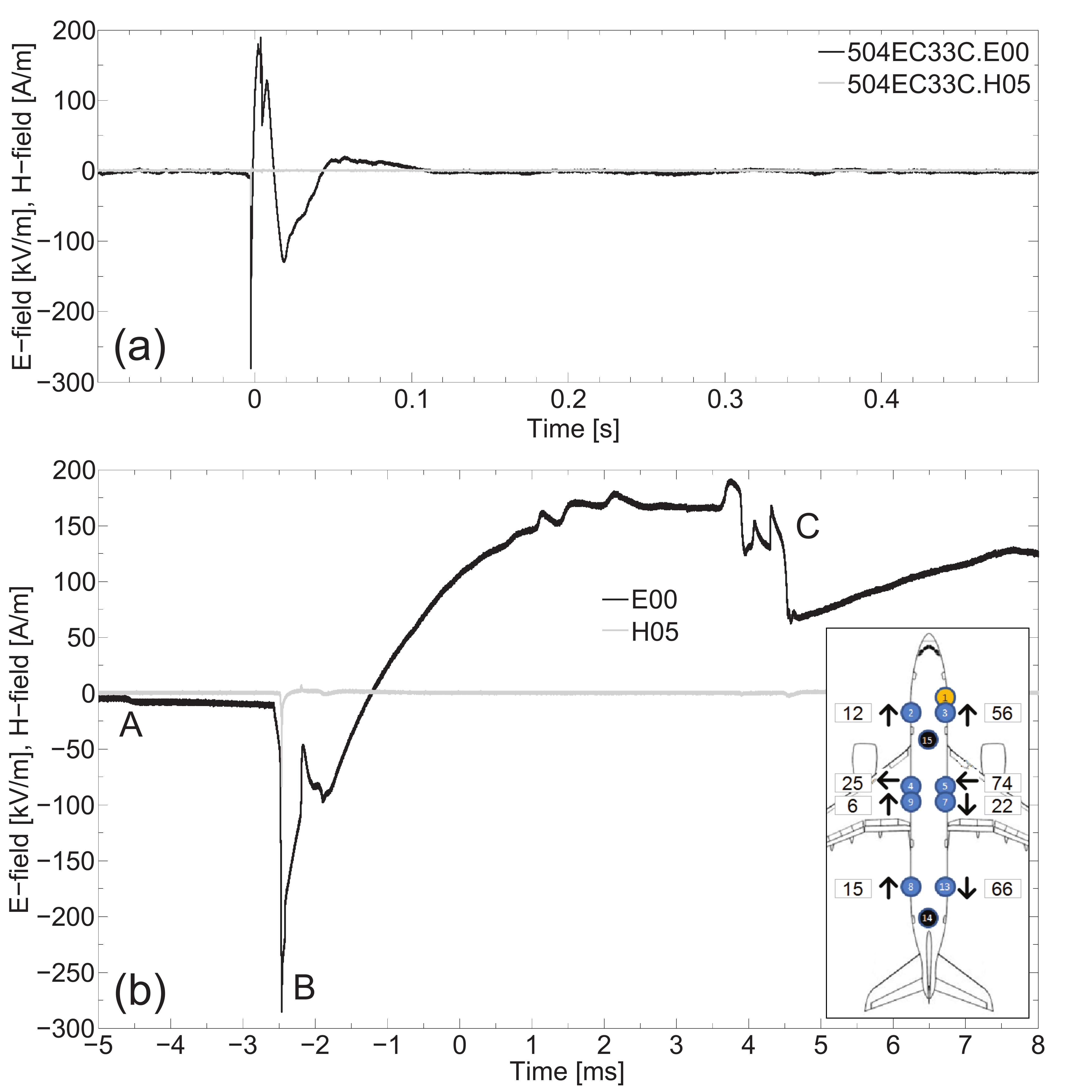}
\caption{The aircraft intercepted lightning. A lightning strike occurred on April 24 at 18:03:52~UTC, our event code 504EC33C. The inset shows the sheet current density distribution at the largest peak. Each arrow in the inset shows the current direction, while the numbers indicate absolute current density in A/m.}
\label{fig:504EC33C}
\end{minipage}
\end{figure}

Figure~\ref{fig:504EC33C}(a) shows 0.5~s of the measurements during the A350 flight on April 24, 2014. This lightning strike occurred at 18:03:52~UTC, our event code 504EC33C. The E and H data come from sensor E00 and H05 respectively. As shown in the zoomed figure~\ref{fig:504EC33C}(b) the E-field start with a small slope at A during approximately 2~ms, followed by an accelerated rise with slope of $-50$~kV/m per $0.1~\mu$s, and then suddenly peaks to $-280$~kV/m. According \cite{Mazur1992a} the sudden change of the E field sensor is indicative for interception of  lightning leader or stroke, rather than the initiation by the aircraft. There is one clear current pulse with most probable current pattern from right wing to nose; see the inset. A small second current pulse occurs near C. Since no return stroke-like current pulses have been observed later, we assume that it was not the main branch of the lightning channel. No x-rays associated with E- or H-signal variations have been observed.

\section{Discussion}
X-rays are generated at two instants, first during the initiation of the lightning channel by and from the aircraft and secondly later in association with recoil processes. At strike initiation the x-ray record was mostly a single photon pulse that could be associated with a step in electric field and a current pulse. This is the moment of first negative stepped leader formation at the aircraft. The leader propagates from the aircraft carrying the X-ray source at its tip \cite{Dwyer2005b}. The second instant occurred about 0.3~s later in both strikes presented. In contrast to the single photon pulses at initiation, the x-rays now come in bursts that last 1 to 4~microseconds during the (re-)attachment of the lightning channel (recoil processes). The bursts have been observed many times, and each time occurred a fraction of a microsecond before the maximum of the E-field. The current maximum occurs then a fraction of the microsecond later. On such time scale the aircraft movement can be neglected. But the recoil streamer head with a speed of 10$^7$~m/s \cite{Mazur1992b} can travel tens of meters.

In a tentative explanation of Figure~\ref{fig:5049BB21}(c), the x-ray source associated with the recoil streamer head approaches the aircraft from the tail. The signal on X14 located aft in the aircraft is correspondingly larger that the signal on X15 between wing box and nose. The current pattern and current reversal indicate that the source moves to the right wing. From the sensor signals alone we cannot distinguish between a jump or a gradual move over the fuselage and wing. The time difference between the rapid termination of the X14 and X15 signals is equal to 350~ns; converted into light-distance this is larger than the size of the aircraft but it suffices to travel several meters for a recoil streamer.

The data in Figure~\ref{fig:203C2BF0} of record 203C2BF0 contain a larger number of recoil processes with many current reversals. Here we only have the X14 data available. Nearly all recoil processes were accompanied by x-rays. The available data confirm the timing with respect to the electrical signals as mentioned for record 5049BB21. It would have been interesting to also have the x-ray record of the three return strokes recorded by ILDAS after $t = 0.4$~s. But here  the SAE design is a compromise between a better time resolution versus a longer record. The current 10~ns per sample was considered to be the longest allowable for our fast x-ray detectors.

We only have one strike 504EC33C out of 61 that can be classified as 'aircraft intercepted', see figure~\ref{fig:504EC33C}(b). At times $t > 0$~ms several steps in the E-field occurred, some of these rounded. According to \cite{Moreau1992}, the steps are due to leader propagation at appreciable distance from the aircraft. Surprisingly, not even a small x-ray signal could be associated with this strike. The source distance should then be larger than the absorption length of 100~keV x-rays in air, or several hundred meters at 4~km altitude.

The rocket triggered lightning experiments \cite{Dwyer2011} revealed that dart leaders generate x-rays during downward propagation that can be detected from distances of several hundred of meters. Also natural stepped leader steps were associated with x-rays, as observed from ground. In this study we observed many recoil processes, most of these associated with the x-rays. It is difficult from the available data to distinguish between dart or dart-stepped leaders, or to discern these from the generic term "recoil process" \cite{Malan1937}.

In individual pulses the absorbed energy in the detector ranges from a few tens of keV to 10~MeV. This agrees with ground based measurement on x-rays in triggered lightning. However, pile-up of several simultaneous lower energy photons cannot be excluded above 2~MeV. In Figure~\ref{fig:5049BB21}(c) the largest energy is about 600~keV.

We fitted the photon energy $E_p$ of the X14 pulses in the burst shown figure \ref{fig:5049BB21}(c) to an exponential probability function $n_p \propto \exp(-E_p/E_c)$ and obtained the characteristic energy $E_c = 170$~keV. The lesser energetic footprint of that event on X15 might be due to the larger distance with respect to the source and/or larger attenuation by air/equipment/fuselage. The brightest x-ray burst occurred in another strike 304FDA20 (not shown here) and was equivalent to $5 \cdot 10^{-12}$~Gy absorbed dose in our detector of 43~cm$^3$ or 0.228~kg.

\section{Conclusions}
New high-resolution well-synchronized data of lightning interaction with an aircraft have been obtained. The magnetic field measurements show that the lightning current direction reverses many times between recoil processes. This feature is a new observation. In contrast to the earlier similar studies \cite{FITZGERALD1967,Pitts1988,Rustan1986,Moreau1992} our measuring system included two x-ray detectors to investigate a lightning- and thundercloud-related hard radiation.

It is shown that lightning strike emits x-rays during its entire lifetime near the aircraft. At the beginning of initiation the x-rays are associated with a negative corona formation on extremities of the aircraft. This observation is consistent with our laboratory studies of long sparks \cite{Kochkin2012,Kochkin2014}. Later, when the aircraft becomes a part of the lightning channel, many recoil processes are accompanied by $\mu$s-fast x-ray bursts. This new observation is important, first because the amount of recoil processes in a single lightning strike to the aircraft is counted by dozens. Secondly, it is shown that a return stroke, being a member of the class of recoil process, can also produce similar x-ray bursts.
We attribute all x-rays to the existence of a high electric field region near the aircraft at the moment of their generation. The field is due to leader or streamers heads; it is strong enough to accelerate electrons to high energies in a so-called thermal run-away breakdown mechanism \cite{Gurevich1961}. The electrons then create x-rays by bremsstrahlung.

Most of the measured lightning strikes are triggered by the aircraft in moderate ambient electric fields of 50~kV/m effectively evacuating charged regions around it. We realize that a fraction of the observed x-rays in the high speed data are probably caused by the presence of the aircraft in the thundercloud. Still, the x-ray observations at 4~km altitude are similar to on-ground measurements in spite of the different air density.

The relation between our data and the natural x-ray occurrence without aircraft requires more investigations. From the magnetic field patterns we conclude that most of our lightning initiations and strokes involve the wings and tail rather that the nose. As an tentative explanation we suggests the engines exhaust plumes with lower density that extend that aircraft imprint on the electric field and form a preferred path for discharges. It is remarkable that the time between the initiation and the first stroke varies between 0.3 and 0.6~s, or with an aircraft displacement between 50 and 100~m, extending the distance streamers and leaders have to span after initiation.

The campaign gave no indication of long gamma-ray glows of a thundercloud. Background changes caused by altitude variations dominated the slowly varying continuous record. A large conductive aircraft campaign might not be the best platform to search for long gamma-ray glows. The aircraft effectively discharges the surrounding volume, thereby collapsing high electric field regions which are possibly responsible for such glows.

Although we did not find a strong and direct evidence of TGFs in our data, we detected several high-energy $\gamma$-ray events while being inside a thundercloud. Further analysis is needed to decide between a cosmic-rays or TGF origin.

\section*{Acknowledgement}
Pavlo Kochkin acknowledges financial support by STW-project 10757, where Stichting Technische Wetenschappen (STW) is part of the Netherlands organization for Scientific Research NWO.

\bibliographystyle{iopams}

\newpage

\begin{figure}[p]
\begin{minipage}{\linewidth}
\includegraphics[width=1\linewidth]{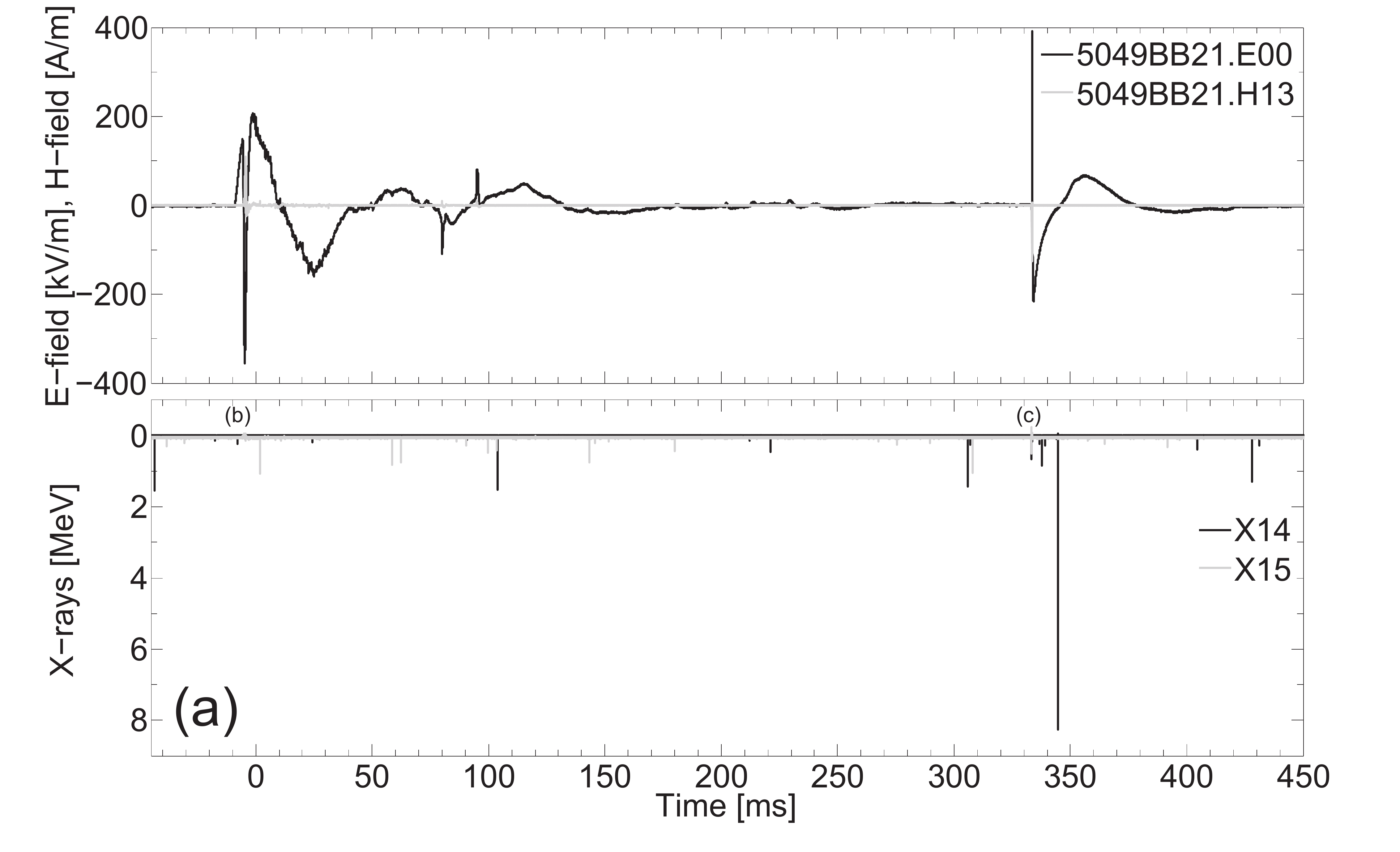}\end{minipage}
\begin{minipage}{\linewidth}
\includegraphics[width=1\linewidth]{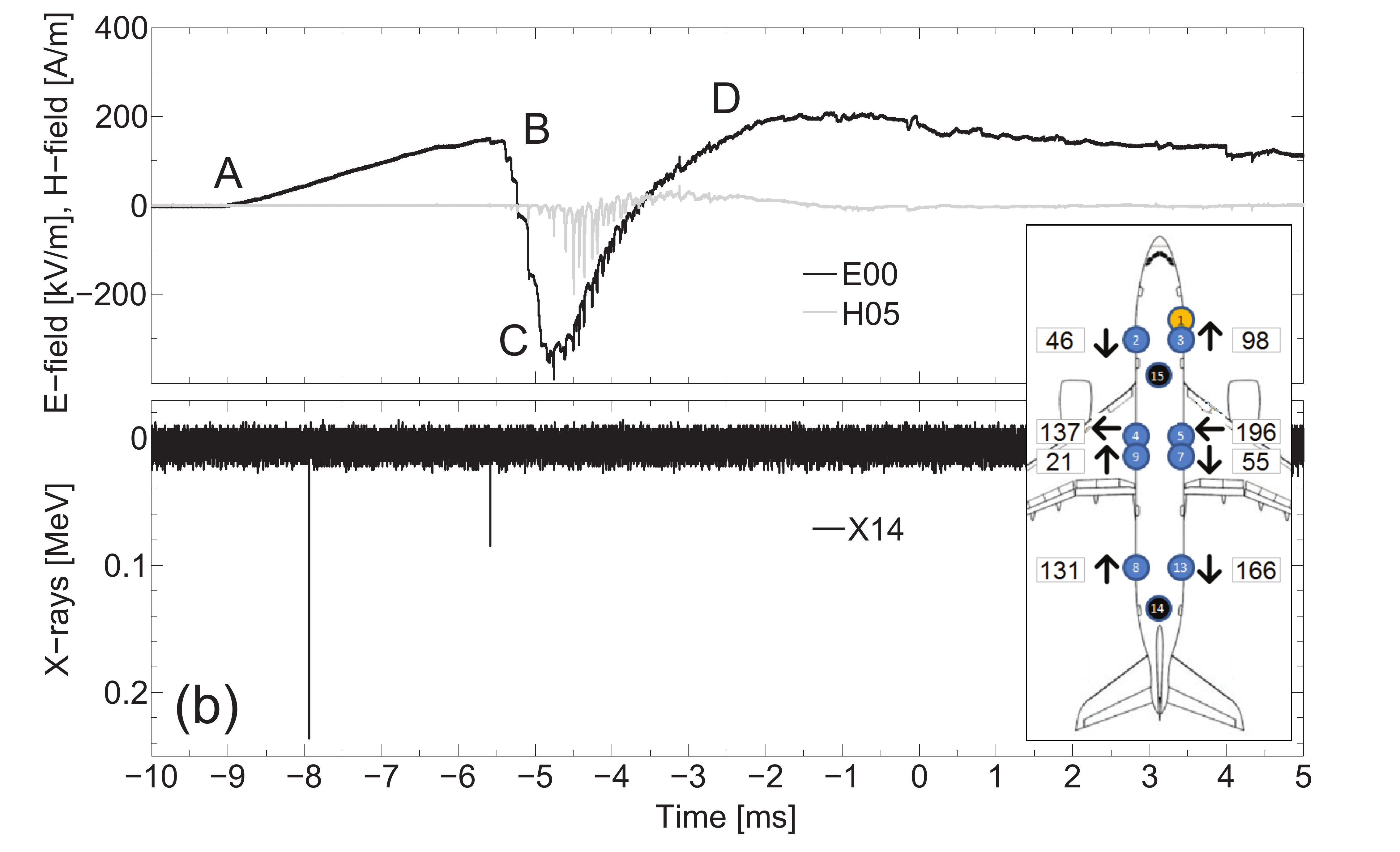}\end{minipage}
\end{figure}
\newpage

\begin{figure}[p]
\begin{minipage}{\linewidth}
\includegraphics[width=\linewidth]{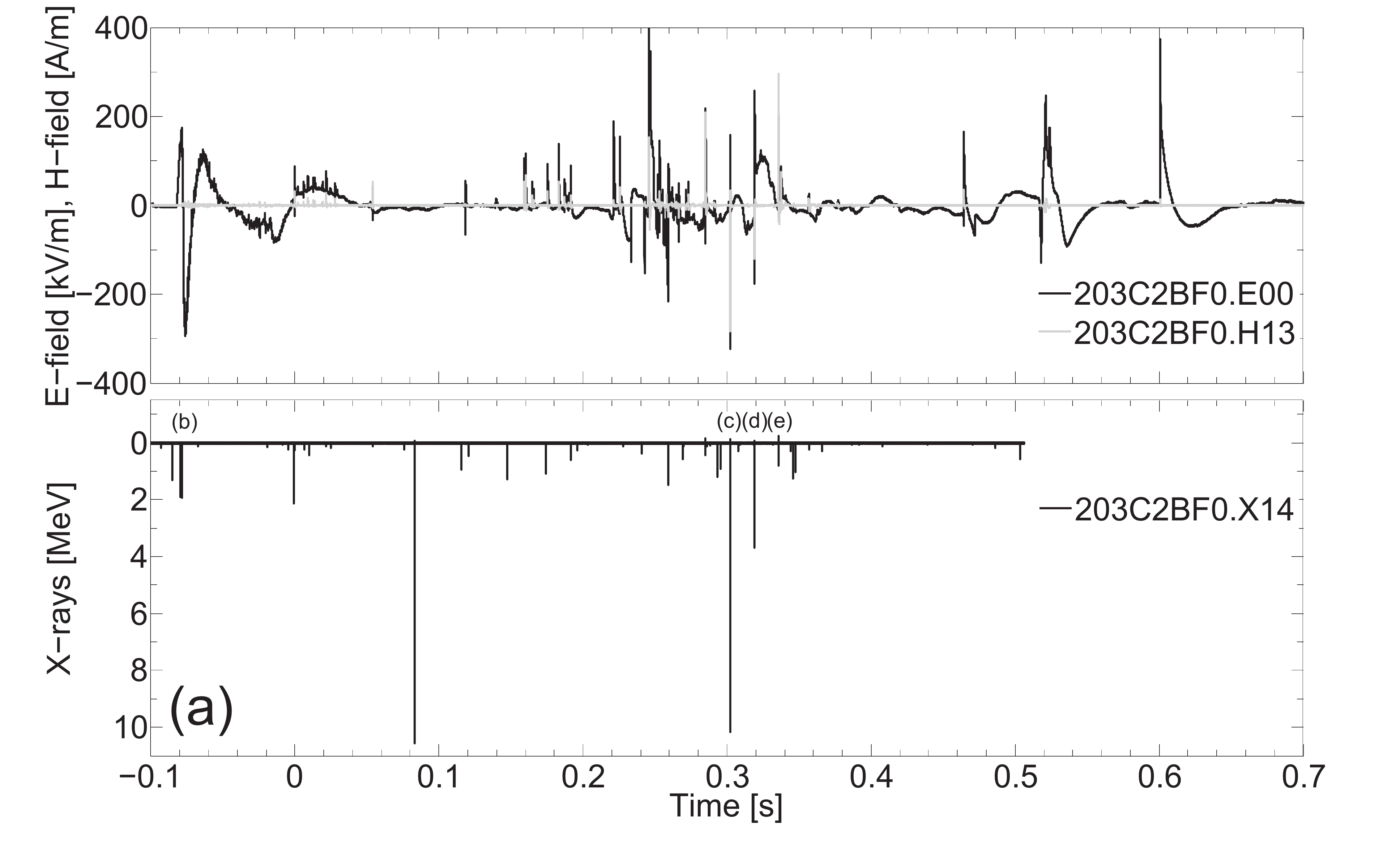}\end{minipage}
\begin{minipage}{\linewidth}
\includegraphics[width=\linewidth]{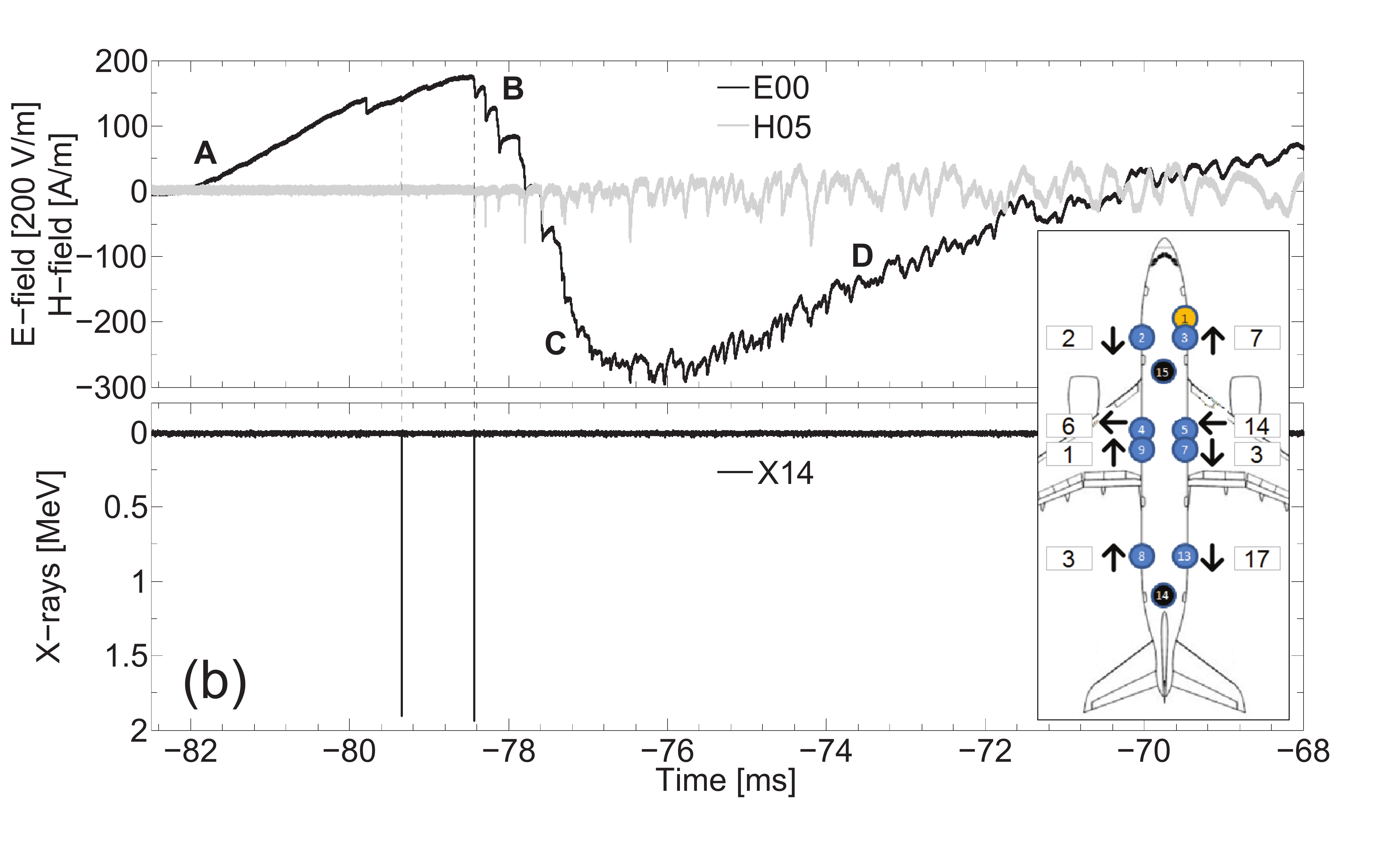}\end{minipage}
\end{figure}
\begin{figure}[p]
\begin{minipage}{\linewidth}
\includegraphics[width=\linewidth]{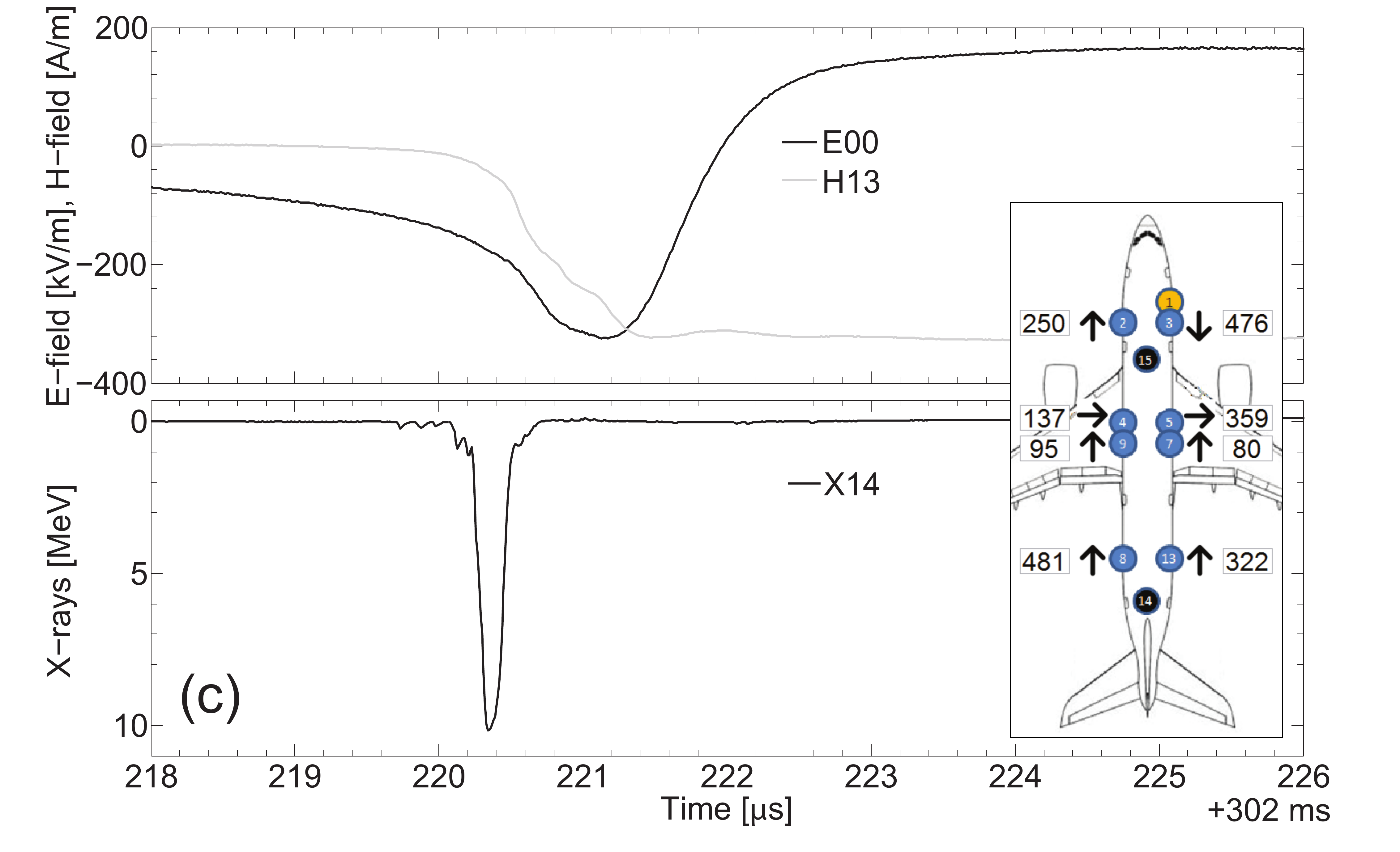}\end{minipage}
\begin{minipage}{\linewidth}
\includegraphics[width=\linewidth]{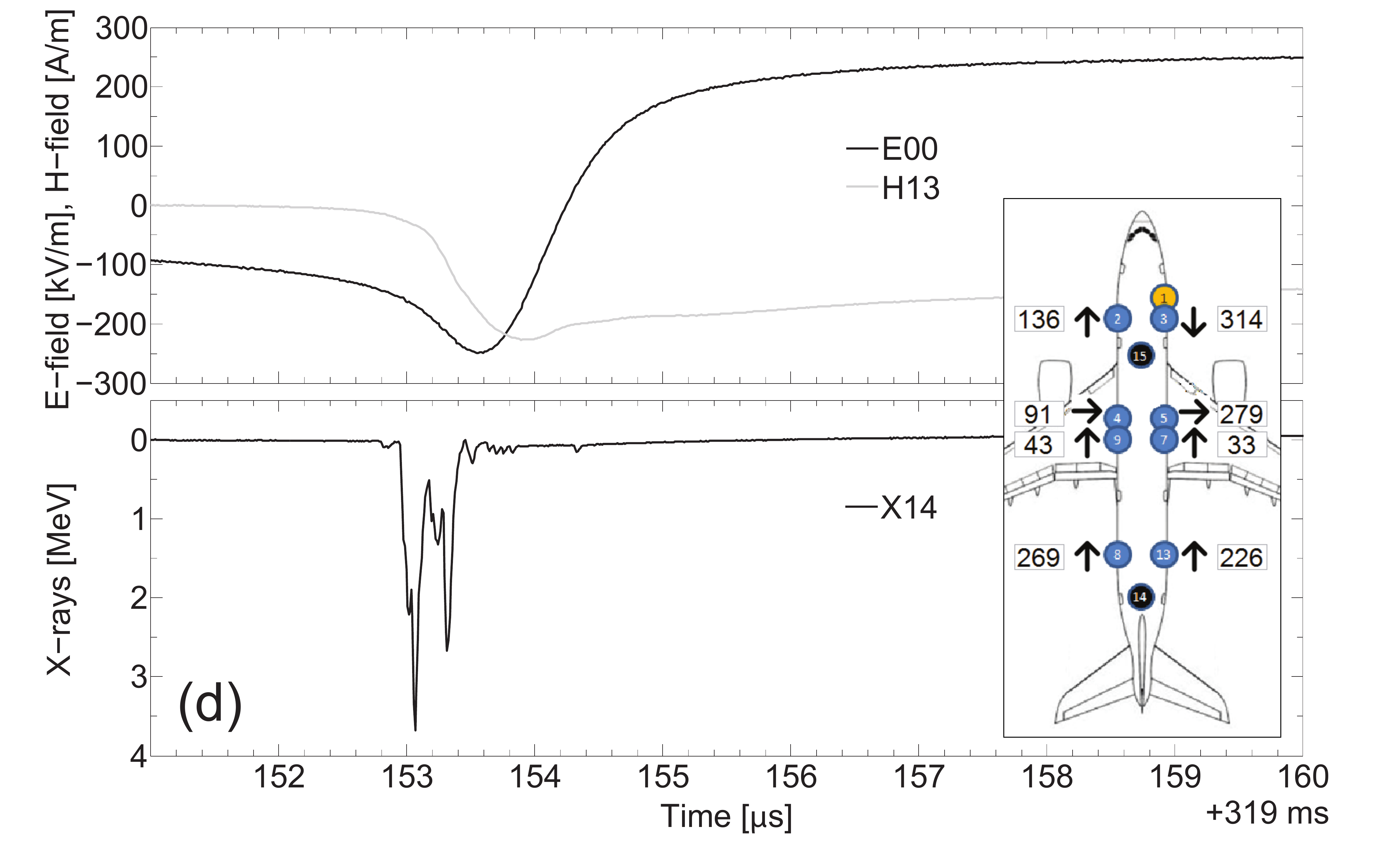}\end{minipage}
\end{figure}

\end{document}